\documentstyle[11pt,newpasp,twoside,epsf]{article}
\def\edcomment#1{\iffalse\marginpar{\raggedright\sl#1\/}\else\relax\fi}
\reversemarginpar

\begin{document}
\title{Variable stars in Stellar Systems}
 \author{Giuseppe Bono}
\affil{INAF - Rome Astronomical Observatory, Via Frascati 33, 
00040 Monte Porzio Catone, Italy; bono@mporzio.astro.it}
 \author{Silvia Petroni}
\affil{Dipartimento di Fisica, Universit\`a degli Studi di
Pisa, Via Buonarroti 2, Pisa, 56127 Italy} 
 \author{Marcella Marconi}
\affil{Osservatorio Astronomico di Capodimonte, Via Moiariello 16, 80131
Napoli, Italy}

\begin{abstract}
We discuss in detail the pulsation properties of variable stars in 
globular clusters (GCs) and in Local Group (LG) dwarf galaxies. Data 
available in the literature strongly support the evidence that we still 
lack a complete census of variable stars in these stellar systems. This 
selection bias is even more severe for small-amplitude variables such 
as Oscillating Blue Stragglers (OBSs) and new exotic groups of variable 
stars located in crowded cluster regions. The same outcome applies to 
large-amplitude, long-period variables as well as to RR Lyrae and 
Anomalous Cepheids in dwarf galaxies.  
\end{abstract}

\section{Introduction}

Variable stars in stellar systems such as GCs and dwarf galaxies have 
played a fundamental role in improving our knowledge 
on stellar populations (Baade 1958) as well as on the physical 
mechanism that drive the pulsation instability (Schwarzschild 1942). 
The main advantage of cluster variables when compared with field ones 
is that they are located at the same distance, and possibly the same 
reddening. Moreover, they formed from the same proto-globular cloud 
and therefore they have the same age, and chemical composition.   
Even though cluster variables present several undoubted advantages 
current knowledge concerning the pulsation properties of these 
objects is still limited. Recent estimates based on new data reduction 
procedures to perform differential photometry (ISIS, Alard 2000) suggest 
that the incompleteness factor in the detection of RR Lyrae stars is at 
least of the order of 30\% (Kaluzny et al. 2001; Corwin \& Carney 2001) in 
Galactic GCs characterized by high central densities.  
This limit is even more severe for OBSs, 
since the luminosity amplitude range from hundredths of a magnitude to a 
few tenths. Moreover, their radial distribution peaks toward the center 
of the cluster, and therefore ground based observations are strongly 
limited by crowding (Gilliland et al. 1998; Santolamazza et al. 2001).  
The same outcome applies to Miras and to Semi-Regular variables in GGCs, 
but for a different reason, quite often they are saturated in current 
CCD chips. This is a real limit for metal-rich clusters of the Galactic 
bulge, since they lack of RR Lyrae stars or host a few of them 
(Pritzl et al. 2002), and the detection of Miras could supply an 
independent distance estimate (Feast et al. 2002).  

Variable stars in dwarf spheroidal (dSph) galaxies presents several pros 
and cons when compared with variables in GGCs. The star formation history 
as well as the dynamical evolution of dSph galaxies is much more 
complex than for GGCs. Typically the age of stellar populations in LG 
dSphs ranges from a few Gyr to 12-13 Gyr, i.e. as old as stars in 
GGCs (Da Costa 1999). Wide photometric surveys strongly support the 
evidence of extra-tidal stars near several dSphs (Irwin \& Hatzidimitriou 
1995; Martinez-Delgado et al. 2001). The observation of 
these stellar debris resembles the tidal tails detected in several GGCs 
(Leon et al. 2000). On the other hand, dSph galaxies apparently host 
large amounts of Dark Matter (DM), and indeed the mass-to-light ratios 
in these systems range from $(M/L)_V\sim 5$ (Fornax) to $\sim100$ 
(Ursa Minor). However, the scenario is still quite controversial and 
the evidence that dSphs present large DM central densities would suggest 
that they are not a large version of GGCs, since the latter present
M/L ratios $\approx1-2$. Photometric and spectroscopic data on variable 
stars in dSphs might supply new insights on the impact that 
environmental effects have on their evolutionary and pulsation properties. 
Unfortunately, data available in the literature are limited, since these 
stellar systems cover wide sky regions. The use of wide field imagers and 
wide field, multifiber spectrographs might overcome these problems.  

In the following we discuss the impact that variables in stellar systems 
might have on cosmic distances and on stellar populations.  
 
\section{Variables in globular clusters}

RR Lyrae stars together with subdwarf main sequence fitting are the 
most popular standard candles to estimate the distance to GGCs 
(Carretta et al. 2000; Bono et al. 2001). Both of them require 
accurate evaluations of cluster metal abundance, but the latter 
ones are more sensitive to reddening corrections (Castellani 1999). 
RR Lyrae stars present the non trivial advantage that individual 
reddening can be estimated on the basis of mean colors. During 
the last few years have been suggested new methods that rely on 
observables that do not depend at all on color excess, namely the 
pulsation period and the luminosity amplitude (Kovacs \& Walker 2001;
Piersimoni et al. 2002). Even though these pulsation parameters 
can be easily estimated, the accuracy of individual reddenings might 
be affected by systematic uncertainties. Empirical evidence suggest 
that approximately the 30\% of fundamental pulsators are affected 
by the Blazhko phenomenon (Kolenberg, this meeting), i.e. the light 
curve shows both amplitude and possibly phase modulation 
(Kurtz et al. 2000). The previous number fraction is supported by 
recent multiband investigation of RR Lyrae in NGC~3201 (Piersimoni 
et al. 2002) and in M3 (Corwin \& Carney 2001). 

\begin{figure}  
\centerline{\epsfxsize= 10.0 cm \epsfbox{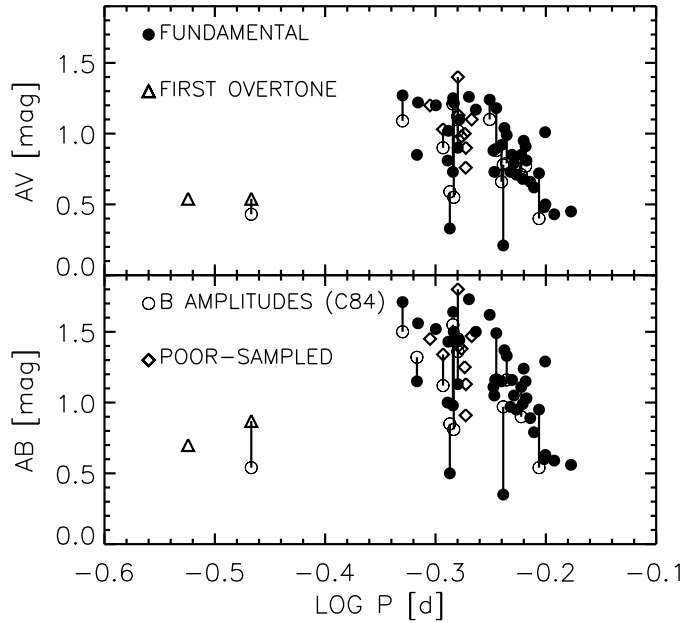}}
\vspace*{0.75 cm}
\caption{Bailey Diagram for RR Lyrae stars in NGC~3201. Filled circles 
and triangles display fundamental and first overtone RR Lyrae. RR Lyrae 
that show amplitude modulations both in the B and in the V band, when 
compared with the amplitudes provided by Cacciari (1984, open circles), 
are connected with a vertical line.} 
\end{figure}

Fig. 1 shows the 
suspected Blazhko RR Lyrae detected in NGC~3201. Note that secondary 
Blazhko periods are only available for a few GGCs such as M3. Although, 
this pulsation feature was detected long time ago (Blazhko 1907) we still 
lack a firm knowledge of the physical mechanisms that drive the occurrence 
of such a phenomenon. Moreover, empirical data for cluster variables 
are poor, since they typically cover short-time intervals. This limits 
the use of the Bailey diagram (amplitude vs period) not only to estimate 
the intrinsic parameters of RR Lyrae (Bono et al. 1997) but also to 
estimate their individual color excesses.  

This limit affects not only the detection of Semiregular (SR) and 
Long-Period-Variables (LPVs) but also variables along the RGB and 
long-period binary systems. On the other hand, the poor spatial 
resolution of ground based measurements and the limited accuracy 
hampered the detection of low-amplitude variables such as SX Phoenicis 
stars and BY Draconis in the innermost regions of GCs. The unprecedented 
amount of homogeneous and accurate 
time series data collected by Gilliland and collaborators to detect 
planets around G type stars in 47~Tuc demonstrated that current knowledge 
concerning cluster variables is still limited. In particular, they found 
a wealth of binary systems, as well as a new class of variable stars 
located at the base of the sub giant branch that they called 
``Red Stragglers'' (see Table 1).  

\begin{table}
\begin{center}  
\hspace{0.25truecm} Table 1. Variable stars detected in 47 Tucanae

\begin{tabular}{lrccc}
               &       &             &          &          \\
\tableline 
Class          &  N$^a$&    $A_V^b$  &Period$^c$&  Source$^d$ \\
               &       &    mag      &    days  &          \\
\tableline 
SRs \& LPVs    &  14   &   \ldots    &  \ldots  &  1         \\  
RR Lyrae       &  1    &  $\approx 1$&  0.738   &  2         \\  
SX Phoenicis   &  6    &  0.01-0.09  & 0.03-0.1 &  3,4       \\  
Det. Ecl. Bin. &  11   &   \ldots    & 0.5-10   &  5         \\  
W UMa          &  15   &   \ldots    & 0.2-0.53 &  5         \\  
Short-Period   &  10   &   \ldots    & 0.1-1.5  &  5         \\  
BY Draconis    &  65   &  0.001-0.04 & 0.5-10   &  5         \\       
CVs            &  9    &  \ldots     & \ldots   &  5         \\
Red Stragglers &  6    &  0.003-0.12 & 1-9      &  5         \\     
Red Giants     &  27   &  \ldots     & 3-10     &  5         \\  
LMXB           &  2    &$\approx0.05$& 0.23-0.36&  6         \\  
MSP            &  20   & 0.004       & 0.43     &  7         \\            
\tableline 
               &       &             &          &          \\
\end{tabular}

\noindent  
$^a$ Number of variables. $^b$ Luminosity amplitude in the V band. 
$^c$ Pulsation period. $^d$ Sources: 1) Fox 1982;  2) Carney et al. 1993;   
3) Gilliland et al. 1998;  4) Bruntt et al. 2001; 
5) Albrow et al. 2001; 6) Edmonds et al. 2002; 
7) Edmonds et al. 2001.  
\end{center}  
\end{table} 

These facts further strengthen the evidence that the knowledge of 
periodic and aperiodic phenomena among cluster stars might be  
biased by selection effects (luminosity amplitudes and time resolution).   
Ground based observations can certainly help to overcome these 
limits for GGCs with low-central densities, but for high-central 
densities and post-core-collapse clusters the use of HST is mandatory.

\section{Variables in dwarf galaxies}

Photometric investigations of variable stars in nearby dwarf galaxies 
have been hampered by the reduced field of view of current CCDs. These 
stellar systems are characterized by low central densities and very 
large tidal radii (Mateo 1998). However, during the last few years 
wide field imagers (WFI) with fields of view of the order of 0.2-0.3 
degree$^2$ become available\footnote{In this site you can find more 
detailed information concerning present and future WFIs 
http://www.ls.eso.org/lasilla/sciops/2p2/E2p2M/WFI}. The absolute 
and the relative calibration of individual CCD chips is quite often 
challenging. Recent results concerning time series data collected 
with the WFI@2.2m ESO/MPI telescope seem to suggest that these 
thorny problems can be overcome at the level of a few hundredths 
of a magnitude (Monelli et al. 2002).  
   
The number of LG dwarf galaxies for which is available 
a detailed census of variable stars is limited (Mateo 1998; 
Cseresnjes 2001; Bersier \& Wood 2002). This limit applies not only to 
long-period and aperiodic variables but also to classical ones 
such as RR Lyrae (Dall'Ora et al. 2002) and $\delta$ Scuti stars
(Mateo, Hurley-Keller, \& Nemec 1998). The reasons why dwarf galaxies 
might play a crucial role in improving current knowledge on stellar 
populations are manifold.  

{\bf i)} Dwarf galaxies harbor stellar populations whose age might 
range from less than 1 Gyr to more than 12 Gyr, i.e. the stellar 
masses range from $M/M_\odot\approx0.1$ to $M/M_\odot\approx2$. 
Therefore they are fundamental laboratories to investigate variable 
stars that are not present in GGCs such as Anomalous 
Cepheids (AC)\footnote{Only one AC is known among GGCs (NGC~5466), see 
Bono et al. (1997) and Corwin, Carney, \& Nifong (1999).}. Recent 
findings based on 
evolutionary and pulsation properties support the evidence that these 
objects are intermediate-mass stars during their central He-burning 
phase (Dall'Ora et al. 2002).  However, we cannot exclude that some 
ACs could be the result of mass transfer in old binary systems 
(Renzini, Mengel, \& Sweigart 1977).  
Moreover, dwarf galaxies might supply fundamental constraints on the 
accuracy of the Period-Luminosity (PL) relation of $\delta$ Scuti. In 
fact, these stellar systems often host both RR Lyrae and $\delta$ Scuti,  
and therefore independent distances may be derived to reduce the 
systematic uncertainties. It is noteworthy that dwarf galaxies, in 
contrast with open clusters, host simultaneously $\delta$ Scuti variables,  
i.e. young intermediate-mass stars, and OBSs, 
i.e. intermediate-mass stars formed via binary collision or binary 
merging of two old, low-mass stars (Santolamazza et al. 2001). 

{\bf ii)} Even though GGCs are the template of low-mass population II 
stars, the HB morphology is affected by the second parameter problem. 
This means that two GGCs with the same metal abundance might have quite 
different stellar distributions on the ZAHB. Dwarf galaxies supply 
the unique opportunity to test whether the dynamical history somehow 
affects the HB morphology.  

{\bf iii)} The number of GGCs that host sizable samples of RR Lyrae 
is limited, while dwarf galaxies with old stellar populations present 
large samples of RR Lyrae. This means that they can be soundly adopted 
to constrain the accuracy of theoretical predictions concerning the 
topology of the instability strip.  

\begin{figure}
\centerline{\epsfxsize= 10.0 cm \epsfbox{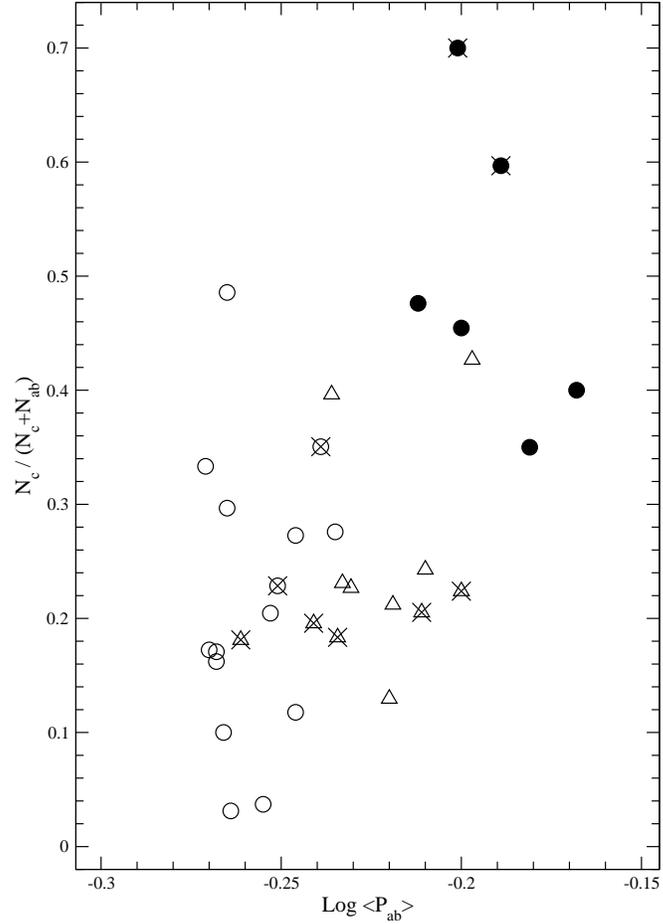}}  
\caption{Ratio between first overtone RR Lyrae ($N_c$) and the total 
number of RR Lyrae versus the mean period of fundamental ($N_{ab}$) 
RR Lyrae. Open and filled circles show Oosterhoff type I and type 
II GGCs that host a sizable sample of RR Lyrae stars. Triangles refer 
to LG dwarf galaxies. Stellar systems in which mixed-mode pulsators 
have been identified are marked with an ``X''. Note that mixed-mode 
pulsators were included among $RR_c$ variables.}  
\end{figure}

To investigate in more detail the last point, Fig. 2 shows the number 
fraction between first overtones and the total number of RR Lyrae as a 
function of the mean fundamental period. Data plotted in this figure 
show that a few dwarf galaxies such as Carina, Draco, and Leo II (see 
data listed in Table 2) can be hardly classified as Oosterhoff type I 
($<P_{ab}>\approx0.55$) or Oosterhoff type II 
($<P_{ab}>\approx0.64$) clusters. They present mean $<P_{ab}>$ values that 
are typical of Oo type II clusters, but the ratio between $RR_c$ and total 
number of RR Lyrae is more typical of Oo type I 
($N_c/(N_{ab}+N_c)\approx0.2$) than of Oo type II 
($N_c/(N_{ab}+N_c)\approx0.5$) clusters. On the other hand, Ursa Minor 
present a mean metallicity quite similar to Draco ($[Fe/H]=-2.2\pm0.1$ vs 
$[Fe/H]=-2.0\pm0.1$, Mateo 1998) but the mean $<P_{ab}>$ value and the 
number ratio of $RR_c$ variables is typical of Oo type II clusters.  
Unfortunately, the number of dwarf galaxies in which have been identified  
mixed-mode variables is still limited and no firm conclusion concerning 
their occurrence can be drawn.  

\begin{table}
\begin{center}  
\hspace{0.25truecm} Table 2. Catalogue of RR Lyrae variables in dwarf galaxies
\begin{tabular}{lccrcrc}
               &       &             &          &          \\
\tableline 
Name &[Fe/H] & $\sigma$ [Fe/H]$^a$ & $N_{ab}$ & $\log \langle P_{ab} \rangle$ &  $N_c$  & Source$^b$\\
\tableline 

Carina      &-2.0$\pm$0.2 &  $<$0.1        &  52 &  -0.200 &  15  &    1 \\
Draco       &-2.0$\pm$0.2 &  0.5$\pm$0.1   & 209 &  -0.211 &  54  &    2 \\
LeoI        &-1.5$\pm$0.4 &  0.3$\pm$0.1   &  47 &  -0.220 &   7  &    3 \\
LeoII       &-1.9$\pm$0.1 &  0.3$\pm$0.1   & 106 &  -0.210 &  34  &    4 \\
Sculptor    & -1.8$\pm$0.1&  0.3$\pm$0.05  & 134 &  -0.236 &  88  &    5 \\
Sextans     &-1.7$\pm$0.2 &  0.2$\pm$0.05  &  26 &  -0.219 &   7  &    6 \\
Ursa Minor  &-2.2$\pm$0.1 &  $\leq$0.2     &  47 &  -0.197 &  35  &    7 \\
Sagittarius &-1.0$\pm$0.2 &  0.5$\pm$0.1   &1906 &  -0.241 & 464  &    8 \\
Fornax      &-1.3$\pm$0.2 &  0.6$\pm$0.1   & 396 &  -0.233 & 119  &    9 \\  
Gal. Center & \ldots      &  \ldots        &1496 &  -0.261 & 331  &    8 \\  
LMC         &-1.7$\pm$0.3 &  \ldots        &3499 &  -0.234 & 786  &   10 \\   
SMC         &-1.7         &  \ldots        &  75 &  -0.231 &  22  &   11 \\ 
\tableline 
               &       &             &          &          \\
\end{tabular}

\noindent  
$^a$ Spread in metallicity;  $^b$ Sources: 1) Dall'Ora et al. 2002; 
2) Nemec 1985; 
3) Held et al. 2001; 4) Siegel \& Majewski 2000; 5) Kaluzny et al. 1995; 
6) Mateo, Fischer \& Krzeminski 1995; 7) Nemec, Wehlau \& Oliveira 1988;
8) Cseresnjes 2001; 9) Bersier \& Wood 2002; 10) Alcock et al. 1996; 
11) Graham 1975; Smith et al. 1992. 
\end{center}  
\end{table}

\section{Discussion}

The results presented in the previous sections bring forward the evidence  
that the empirical scenario for variable stars in stellar systems such as  
GCs and LG dwarf galaxies is far from being complete. The limit applies 
not only to aperiodic variables but also to long-period variables such as 
Miras and Semi-Regulars. The same outcome applies to RR Lyrae stars affected 
by the Blazhko effect.  

Even though several LG dSphs present stellar populations with 
chemical compositions and stellar ages quite similar to stars in GCs, 
the RR Lyrae variables present pulsation properties that are a {\em bridge} 
between Oosterhoff type I and Oosterhoff type II clusters. This preliminary 
evidence seems to suggest that either the dynamical history and/or the 
chemical evolution in these stellar systems might play a role to explain 
this difference. In this context the use of the new wide field imagers 
will supply the unique opportunity to investigate on a star-by-star basis 
the stars and the variables in LG dwarf galaxies.  
 
Although new theoretical frameworks have been developed to account for 
the occurrence of mixed-mode pulsators among RR Lyrae and OBSs we still 
lack a comprehensive explanation of the 
physical mechanisms that drive the occurrence of such a phenomena.  
It goes without saying that new sets of full amplitude nonlinear, 
convective models tightly connected with evolutionary models are 
highly requested to constrain the region of the instability strip 
where these pulsators present this intriguing behavior.  


\end{document}